# Orbital angular momentum complex spectrum analyzer for vortex light based on rotational Doppler Effect

## Running title: OAM complex spectrum analyzer

Hailong Zhou,[1+] Dongzhi Fu,[2+] Jianji Dong,[1,*] Pei Zhang,[2,*] Dongxu Chen,[2] Xinlun Cai,[3] Fuli Li,[2] and Xinliang Zhang[1]

[+]These authors contributed equally to this work

*Correspondence and requests for materials should be addressed to Jianji Dong (email: jjdong@hust.edu.cn), and Pei Zhang (zhangpei@mail.ustc.edu.cn).

**Contact details of all authors**. Jianji Dong: Email: jjdong@mail.hust.edu.cn, Phone: 86-27-87792367, Fax: 86-27-87792367; Hailong Zhou, zhouhailongshe@126.com; Dongzhi Fu, fdz1010@stu.xjtu.edu.cn; Pei Zhang, zhangpei@mail.ustc.edu.cn; Dongxu Chen, xdc.81@stu.xjtu.edu.cn; Xinlun Cai, caixinlun@gmail.com; Fuli Li, flli@mail.xjtu.edu.cn; Xinliang Zhang, xlzhang@mail.hust.edu.cn.

[1]Wuhan National Laboratory for Optoelectronics, School of Optical and Electronic Information, Huazhong University of Science and Technology, Wuhan, China, 430074

[2]Key Laboratory for Quantum Information and Quantum Optoelectronic Devices, Shaanxi Province, Department of Applied Physics, Xi'an Jiaotong University, Xi'an 710049, China

[3]State Key Laboratory of Optoelectronic Materials and Technologies and School of Physics and Engineering, Sun Yatsen University, Guangzhou 510275, China.

**Abstract: The function to measure orbital angular momentum (OAM) distribution of vortex light is essential for OAM applications. Although there are lots of works to measure OAM modes, it is difficult to measure the power distribution of different OAM modes quantitatively and instantaneously, let alone measure the phase distribution among them. In this work, we demonstrate an OAM complex spectrum analyzer, which enables to measure the power and phase distribution of OAM modes simultaneously by employing rotational Doppler Effect. The original OAM mode distribution is mapped to electrical spectrum of beating signals with a photodetector. The power distribution and phase distribution of superimposed OAM beams are successfully retrieved by analyzing the electrical spectrum. We also extend the measurement to other spatial modes, such as linear polarization modes. These results represent a new landmark of spatial mode analysis and show great potentials in optical communication and OAM quantum state tomography.**

# INTRODUCTION

Vortex light carries orbital angular momentum (OAM) characterized by $\exp(il\theta)$, where $\theta$ is the angular coordinate and $l$ is the topological charge (TC)[1]. These OAM beams have been widely used in a variety of interesting applications, such as micromanipulation[2,3], probing the angular velocity of spinning microparticles or objects[4,5], quantum information[6,7] and optical communication[8,9]. Apparently, TC is a basic physical parameter to characterize OAM light. The capability of distinguishing different OAM modes is essential in OAM-based optical system. There are many interference methods to covert the OAM modes into identifiable intensity patterns, such as holographic detection with plasmonic photodiodes[10] and diffraction patterns of various apertures[11-15]. In these schemes, the hybrid OAM modes are hard to distinguish. To distinguish the hybrid OAM states, a common approach is to convert the unknown OAM modes to the fundamental mode (TC=0) with a spatial light modulator (SLM) and then calculate the power ratio of each mode after mode filtering[16-18]. This approach needs to measure the power of OAM modes one by one, or requires multiple photodetectors, which is either time-consuming or cumbersome to use. Another popular approach is to sort the hybrid OAM modes into different spatial locations, such as Cartesian to log-polar coordinate transformation[19-22], interferometric methods based on a rotation device[23-25]. These techniques require photodetector arrays to detect the separated OAM states and it is difficult to measure a large number of hybrid OAM modes for the finite diffraction space. There are also some progresses of measurement of the power spectrum by mapping the OAM spectrum to time[26,27], but the measured range and phase detection are limited. Moreover, it cannot measure the OAM modes with negative TCs. In view of the imperfect measurement of state of the art, it is desirable to develop an OAM complex spectrum analyzer enabling instantaneous and accurate measurement of OAM mode distribution of light, similar to the function of optical spectrum analyzer to characterize the frequency distribution. The OAM complex spectrum analyzer is defined as a device that can measure the power and the phase distributions of OAM components simultaneously. In 2014, the complex probability amplitudes of OAM states were successfully measured at the single photon level through sequential weak and strong measurements[28]. However, OAM complex spectrum analyzer in classical information system has been rarely reported and achieved yet. In recent years, a new transversal Doppler effect (rotational Doppler Effect) was demonstrated associated to the transverse helical phase and the Doppler velocimetry of rotating objects was developed based on this effect[4,29]. In turn, this effect shows great potential for OAM complex spectrum analyzer.

In this article, we demonstrate a prototype of OAM complex spectrum analyzer by employing rotational Doppler Effect. The system consists of unknown input OAM beams, a strong reference light, a spinning reflective object, a mode filter and a photodetector. The original OAM mode distribution is mapped to electrical spectrum of beating signals detected by a photodetector. The OAM power spectrum is measured with inherent noise via initial measurement. However, the inherent noise can be fully eliminated via subtraction of the initial measurement and another constant-noise measurement. System calibration is also discussed when the input modes are mapped into the beating signals with an unequal efficiency, meanwhile, the phase distribution of OAM modes can also be acquired via calibration. Similar to the optical spectrum analyzer, our scheme has huge potentials in OAM complex spectral analysis and measurement for OAM-based communication systems and quantum systems.

## MATERIALS AND METHODS

When an OAM light illuminates a spinning object with a rotating speed at $\Omega$, as shown in Fig. 1, the scattered light will have a frequency shift, which is related to the change of TC. The reduced Doppler shift is given by[4]

$$\Delta f = (l - m)\Omega / 2\pi \text{ , } \textbf{(1)}$$

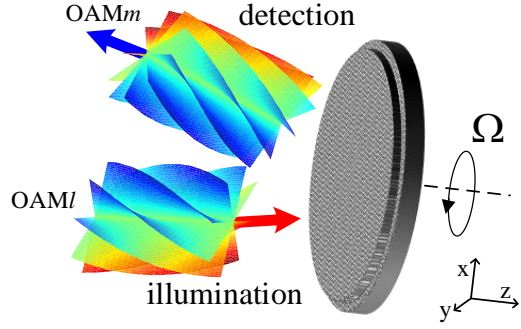

Figure 1. Schematic diagram of rotational Doppler Effect. When an OAM mode illuminates a spinning object, the scattered light will have a frequency shift.

where $l$ is the TC of incident light, $m$ is the TC of scattered light. Equation (1) suggests us that the reduced frequency shift is dependent on the TCs of the incident and scattered light, and the rotating speed. These features can be utilized to measure the OAM complex spectrum of light. In our recent work[30], we presented a model to sufficiently investigate the optical rotational Doppler Effect based on modal expansion method. Here, we provide a simple derivation based on this model to explain how the effect can be used to measure the OAM complex spectrum. For simplicity, we ignore the difference along the radial, so the modulated function of spinning object can be expressed in Fourier expansion form

$$M(r,\theta) = \sum A_n \exp(in\theta)\exp(-in\Omega t) \text{ , (2)}$$

where $n$ are integers and $t$ denotes the time. Assume that the input light consists of lots of unknown OAM modes expressed as $\sum_{s=1,2,\cdots,N} B_s \exp(-i2\pi f t)\exp(il_s\theta)$, whose TCs range from $l_1$ to $l_N$. $f$ is the frequency of light and $B_s$ is the complex amplitude of corresponding mode. We introduce a reference light expressed by $\gamma B_0 \exp(-i2\pi f t)\exp(il_0\theta)$ to illuminate the spinning object with the input light together, where $\gamma$ is a parameter to tune the power of reference light. The scattered light can be deduced as

$$E_o = \sum_{m \in Z} \begin{Bmatrix} \gamma B_0 \exp(-i2\pi ft) A_{m-l_0} \exp(im\theta) \exp[-i(m-l_0)\Omega t] + \\ B_1 \exp(-i2\pi ft) A_{m-l_1} \exp(im\theta) \exp[-i(m-l_1)\Omega t] + \\ B_2 \exp(-i2\pi ft) A_{m-l_2} \exp(im\theta) \exp[-i(m-l_2)\Omega t] + \\ \cdots \\ B_N \exp(-i2\pi ft) A_{m-l_N} \exp(im\theta) \exp[-i(m-l_N)\Omega t] \end{Bmatrix} . \textbf{(3)}$$

After transmitting for a certain distance, only one OAM mode (assume OAM*m*) is selected by mode filtering and then is collected by a photodetector. The collected intensity dependent on the time can be derived as

$$
\begin{aligned}
I_m(\gamma) = & \sum_{s=1,2,\cdots,N} \left| B_s A_{m-l_s} \right|^2 + \left| \gamma B_0 A_{m-l_0} \right|^2 \\
& + 2 \sum_{1 \le p < q \le N} \left| B_p A_{m-l_p} B_q A_{m-l_q} \right| \cos\left[ (l_p - l_q)\Omega t + \phi_{p,q} \right] \\
& + 2\gamma \sum_{1 \le p \le N} \left| B_p A_{m-l_p} B_0 A_{m-l_0} \right| \cos\left[ (l_p - l_0)\Omega t + \phi_{p,0} \right]
\end{aligned}
\textbf{, (4)}
$$

where $\phi_{p,q} = \text{angle}\left( B_p A_{m-l_p} B_q^* A_{m-l_q}^* \right)$. If the power of reference light is much higher than the one of input light, the alternating current (AC) terms of intensity can be approximately expressed as

$$
\begin{aligned}
I_{m,AC}(\gamma) = & 2 \sum_{1 \le p < q \le N} \left| B_p A_{m-l_p} B_q A_{m-l_q} \right| \cos\left[ (l_p - l_q)\Omega t + \phi_{p,q} \right] \\
& + 2\gamma \sum_{1 \le p \le N} \left| B_p A_{m-l_p} B_0 A_{m-l_0} \right| \cos\left[ (l_p - l_0)\Omega t + \phi_{p,0} \right] \\
\approx & 2\gamma \sum_{1 \le p \le N} \left| B_p A_{m-l_p} B_0 A_{m-l_0} \right| \cos\left[ (l_p - l_0)\Omega t + \phi_{p,0} \right]
\end{aligned}
\textbf{. (5)}
$$

If $l_p - l_0 (p = 1, 2, \cdots, N)$ does not change its sign, the OAM modes are one-to-one mapped into the frequencies and the coefficients are related to the corresponding complex amplitudes, so it can be designed as a kind of OAM mode analyzer. If all the coefficients $A_n$ and $B_0$ are known, the OAM complex spectrum can be obtained through spectrum analysis. Note that $\phi_{p,0}$ is a value calculated by a complex integral when considering the difference along the radial, which is not equal to the phase of input OAM mode. Luckily, $\phi_{p,0}$ always has a fixed deviation relative to the one of input OAM mode and the deviation can be calibrated by a pre-measurement, so the phase distribution of input OAM modes can be also retrieved by system calibration.

Note that the cross terms of input OAM modes are ignored, so some noise is introduced in this scheme. In fact, the noise from the cross terms can be removed by twice measurements, only if we change the power of reference light. The subtraction of AC signal of twice measurements is given by

$$I_D = I_{m,AC}(\gamma_2) - I_{m,AC}(\gamma)$$
$$= 2(\gamma_2 - \gamma) \sum_{1 \le p \le N} \left| B_p A_{m-l_p} B_0 A_{m-l_0} \right| \cos\left[ (l_p - l_0)\Omega t + \phi_{p,0} \right]$$

.  (6)

Therefore, by a measurement with a strong reference light (defined as initial measurement), we can measure the OAM power spectrum with inherent noise. And through subtraction of the initial measurement and another measurement by changing the power of reference light (defined as constant-noise measurement), the inherent noise can be eliminated perfectly.

In fact, the characteristic of spinning object is hard to precisely determine in practical applications, so the system calibration is required. The calibration procedure is divided into three steps. Firstly, we fix the power of reference light and input light, and then scan the input OAM mode to measure the complex amplitude (including amplitude and phase) of beating signal between every input mode and the reference mode one by one. The measured complex amplitudes are the reference data for calibration. Secondly, we set the input light as the one to be measured and then measure the complex amplitudes of every harmonic signal according to Eq. (5) or Eq. (6). Finally, the complex amplitude distribution of input OAM modes can be accurately retrieved via dividing the complex amplitudes of every harmonic signal by the reference data.

To evaluate the measurement accuracy, we define the measurement error by

$$\delta = \sqrt{\sum_{s=1}^{N} \left( \left| B_s \right|^2 / \sum_{q=1}^{N} \left| B_q \right|^2 - \left| C_s \right|^2 / \sum_{q=1}^{N} \left| C_q \right|^2 \right)^2 / N}$$

,  (7)

where $\{ B_s \mid s = 1, 2, \cdots, N \}$ are the theoretical OAM complex spectrum and $\{ C_s \mid s = 1, 2, \cdots, N \}$ are the experimental ones.

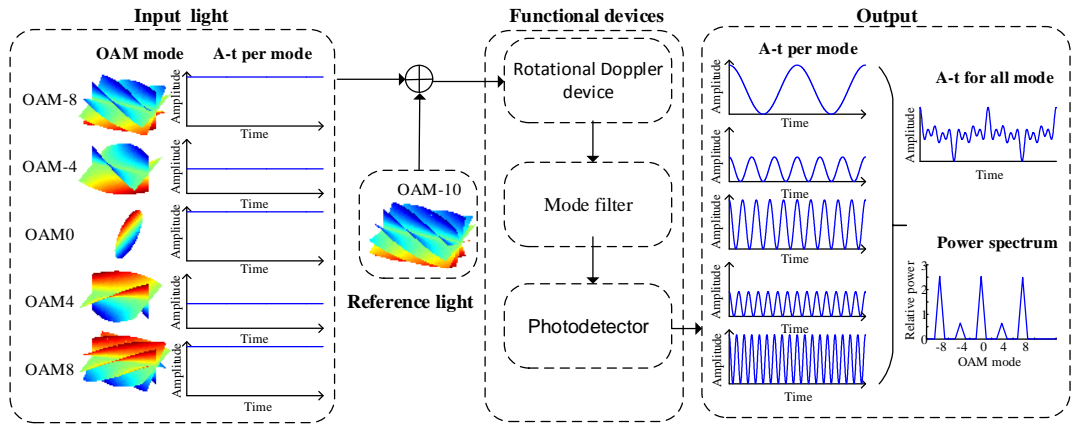

Figure 2. Diagram of OAM complex spectrum analyzer. Input light: The three dimensional color images indicate spiral phase structures of OAM modes and A-t curves present the amplitude dependency on time. Reference light: the reference light is selected as an OAM mode outside the range of input light. Functional devices: the devices contain three parts. The first part is to realize Doppler shift based on rotational Doppler Effect and the mode filter is used to select a certain OAM mode, in general, it is the OAM0. So every mode is converted into OAM0 and has a frequency shift linear with the TC. Every input mode will beat with the strong reference mode resulting to a beating signal and the beating frequency is also linear with the TC. The photodetector is employed to measure the beating signal. Output: the first column presents the beating signals of the reference mode and corresponding input modes. The beating frequency is linear with the TC of input mode and the amplitude is proportional to the one of input mode. The total intensity is chaotic, but the power spectrum, which can be obtained through Fourier transform of the total intensity collected by the photodetector, is consistent with the input OAM power spectrum.

Figure 2(a) shows the diagram of OAM complex spectrum analyzer, which consists of input light to be measured, a strong reference light, a rotational Doppler device, a mode filter and a photodetector. The input light contains many unknown OAM modes within a certain finite-dimensional Hilbert space. For simplicity of description, we assume that the TCs of input OAM modes are within the range from -8 to 8. And we introduce a strong coaxial reference light whose TC is outside the range of input OAM light. Here, we assume that the reference light is OAM-10 (TC=-10). When the input modes and the reference mode simultaneously illuminate a spinning object, they will experience different Doppler frequency shifts. The Doppler frequency shift is linear with the input TC because only the fundamental mode (OAM0) is selected by the mode filter. The frequency differences will result in various beating signals which can be detected by a photodetector. The intensity of reference light is much stronger than that of the input light, so the beating signals among the input modes can be ignored (noise at real measurement) and we only need to consider the beating signals between the input light and the reference light. Shown as the amplitude-time (A-t) curves in Fig. 2, the OAM-8 and the reference mode OAM-10 will result in a beating signal at two times of the rotating speed. Similarly, the frequencies of other beating signal are proportional to the TC differences between the input modes and the reference mode. Thus, the input OAM modes are successfully mapped to the beating frequencies and the amplitude of beating signal is proportional to the one of corresponding input OAM mode. Finally, we can get the OAM power spectrum through Fourier transform of the total intensity collected by the photodetector. It should be noted that we from the beating signals among the input modes, so it is feasible only when the reference light is much stronger than the input light in the initial measurement. The noise is irrelevant to reference light, thus it can be eliminated via subtraction of the initial measurement and another constant-noise measurement. The characteristic of spinning object is hard to determine in practical applications, so the calibration is required and the phase distribution of OAM modes can be also acquired via system calibration. We should notice that since only the OAM0 is collected by the photodetector, the overall efficiency of the scheme is quite low. Even so, it does not affect the measurement accuracy since a high sensitivity photodetector can be used.

# RESULTS AND DISCUSSION

### Experimental Setup

We design a proof of principle experiment to demonstrate the scheme. Figure 3 shows the experimental setup. The light emitted from the He–Ne laser (wavelength at 633 nm) is expanded with two lenses (L1 and L2), and then illuminates SLM1. The half wave plate (HWP) and polarization beam splitter (PBS) are used to select the horizontal polarization to match the operating polarization of SLM and to tune the input power. SLM1 is divided into three parts, shown as the inset. The outer part is fixed and used to generate the input OAM modes to be detected, the middle part is used to introduce the reference light, and the inner part is used to tune the power of reference light via changing the innermost radius. Only the input light (outer part) and the reference light (middle part) are diffracted from the first-order by adding a grating in the two areas. The first-order diffracted beam is selected with a pinhole (P1) and then illuminates the spinning object (SLM2). The pattern of SLM2 is rotated to scan the azimuth. Afterwards, the fundamental mode (OAM0) is selected by another pinhole (P2) and then a charge-coupled device (CCD) is used to collect the light intensity. The lenses (L3, L4, and L5) are employed to tune the optical route.

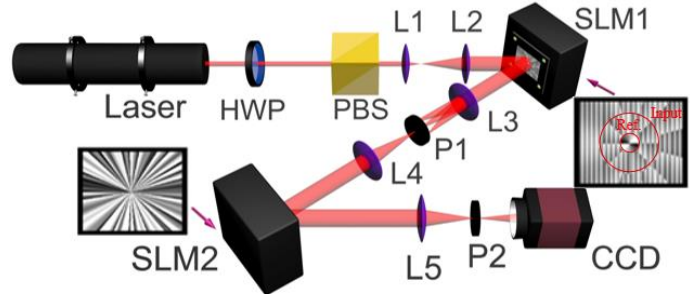

Figure 3. Experimental setup for measurement of OAM complex spectrum. The light emitted from the laser illuminates the SLM1 to generate the input light and the reference light, and then is selected from the first-order diffraction by a pinhole (P1). After that, the light is incident on the spinning object (SLM2) and then the OAM0 is selected from the scattered light by the pinhole (P2). Finally, a charge-coupled device (CCD) is used to calculate the total light intensity of OAM0.

## Measurement of OAM modes

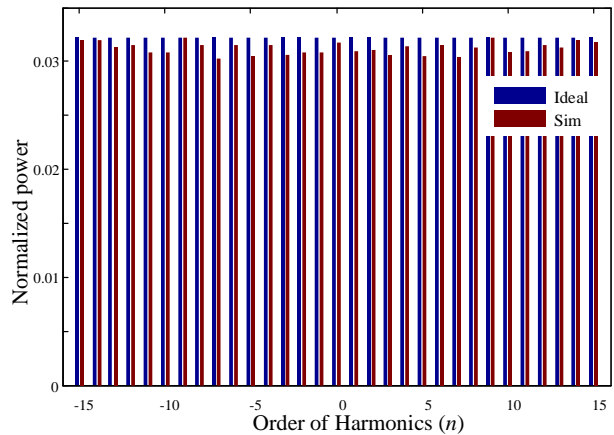

Figure 4. Fourier harmonic distributions of the spinning object. The blue bars denote the ideal distribution and the red bars denote the theoretical distribution by using iterative algorithm.

In order to implement a well-proportioned mapping from OAM modes to electrical spectrum, the spinning object characterized by Eq. (2) should have a uniform power distribution for each helical phase function and it can be achieved with a phase-only element by using iterative algorithm[31,32]. Here, an SLM (SLM2) is employed to emulate the spinning object. Assume that the TCs of input OAM modes are limited in the range from -8 to 8 and the reference light is selected as OAM-10. Thus for the spinning object, an even Fourier expansion of modulation function ranging from -15 to 15 meets the requirement, as shown in Fig. 4. The blue bars denote the ideal power weight of each helical phase component and the red bars denote the theoretical results by using iterative algorithm. We can see that an approximate even Fourier expansion ranging from -15 to 15 is achieved.

We firstly measure some patterns of different OAM states. Figures 5(a) and 5(b) show the simulated far-field patterns and experimental results of OAM-4, OAM-4 mixed with OAM6, OAM-4 mixed with OAM4, and a Gaussian distribution of OAM modes respectively. The experimental patterns are consistent with the simulated ones. In our experiment, we set the OAM mode distribution of input light as the same Gaussian distribution, shown as the blue bars in Fig. 5(h). In the initial measurement, we make the power of reference light larger than the power of input light by tuning the radius of inner part of SLM1. The received intensity by CCD as a function of rotating angle is shown in Fig. 5(c). We then calculate the power spectrum through Fourier transform of the periodic signal collected by CCD. The power spectrum is shown in Fig. 5(d). We can see that the measured OAM power spectrum is roughly consistent with the theoretical predictions but still shows significant deviation, especially in small mode region. The measurement error $\delta$ is equal to 0.0767. The errors are mainly from the neglected beating signals among the input modes, namely, the cross terms in Eq. (5). According to our theoretical model, the cross terms can be suppressed by increasing the power of reference light. In the constant-noise measurement, we increase the power of reference light by reducing the radius of inner part of SLM1 to zero, the received intensity and the corresponding power spectrum are shown in Figs. 5(e) and 5(f) respectively. The measurement error reduces to 0.0366. Apparently, the spectrum deviation becomes smaller, especially in small mode region. This finding proves that the noise from the cross terms in Eq. (5) can be ignored when the power of reference light is much higher than the power of input light. More importantly, the cross terms can be completely removed by subtraction of the initial measurement and the constant-noise measurement in theory as shown in Eq. (6). The measured power spectrum of the intensity subtraction is shown in Fig. 5(g). As expected, the measurement error rapidly diminishes to 0.017 and the noise is greatly suppressed. The final experimental results and the theoretical results are presented in Fig. 5(h), one can see that the experimental results agree well with the theoretical calculations.

Although the measurement power spectrum shows a good consistency with the expected distribution, we still find some deviations in Fig. 5(h), which is mainly caused by the unequal conversion efficiency from the input modes to the beating signals. Meanwhile, it is hard to ensure even conversion efficiency for each mode in practical applications, and the phase distribution cannot be determined owing to the unclear phase mapping, so the calibration is necessary. Through system calibration, we can acquire the complex amplitude (containing amplitude and phase) distribution of OAM modes precisely.

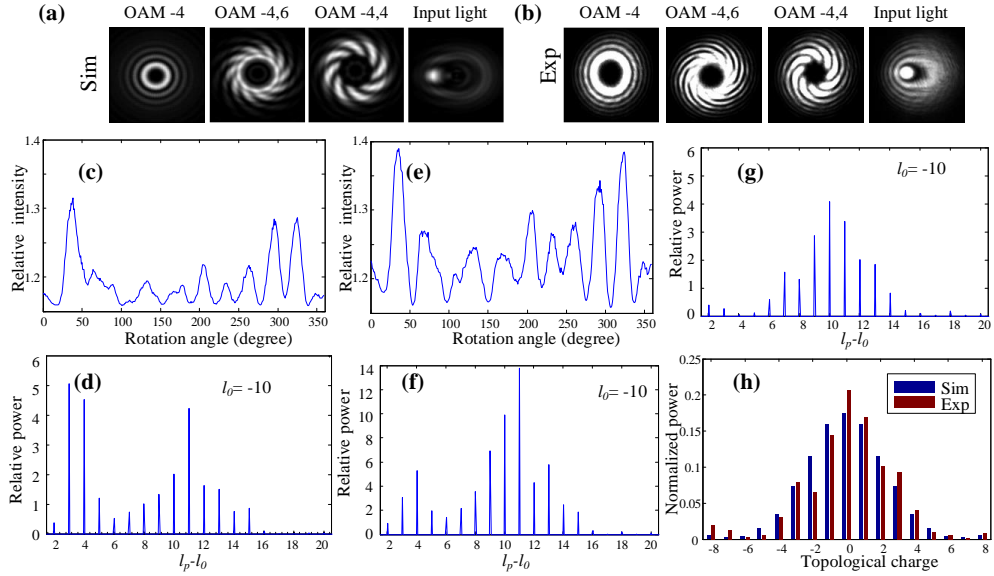

Figure 5. Experimental data for measurement of a Gaussian distribution in the OAM basis. (a, b) The (a) simulated patterns and (b) experimental results of OAM-4, OAM-4 mixed with OAM6, OAM-4 mixed with OAM4, and a Gaussian distribution of OAM modes respectively. (c, d) The measured intensity in one period and the calculated OAM power spectrum by introducing a reference light with TC =-10, in the case, the noise from the neglected cross terms is obvious. (e, f) The results when increasing the power of reference light, it proves that the cross terms can be suppressed by increasing the power of reference light. (g) The OAM power spectrum of twice measurements. The power spectrum of twice measurements has lowest noise. (h) The final experimental results compared to the theoretical results.

**System calibration for a general spinning object**

Now, we consider a more general case where the characteristic of spinning object is nonuniform or unknown. In the case, the conversion efficiency of every OAM mode to OAM0 is inconsonant, so we need firstly get the complex amplitudes of beating signal between every input OAM mode with the reference mode under the same condition. Assume that the TCs of input OAM modes are limited in the range from -10 to 10 and the reference light is selected as OAM-15. Firstly, we keep the power of input light and reference light constant. Then we scan the input OAM mode from OAM-10 to OAM10 and get the corresponding complex amplitudes of beating signal. The amplitude distribution and the phase distribution of beating signal between the input OAM modes and the reference mode are shown in Figs. 6(a) and (b) respectively. One can see that complex amplitude distribution is messy, it is because that the modulation function of spinning object has a nonuniform harmonic distribution and the phase mapping is unclear. Here, the phase distribution in Fig. 6(b) is exactly the fixed deviation between $\phi_{p,0}$ and the phase distribution of input OAM modes. After getting the complex amplitude, we can start to measure the OAM complex spectrum. For example, we design a simple OAM mode superposition of input light, which is to be measured. The input modes consist of finite modes (TC=-10, -5, 0, 5, 10), whose initial intensity and phase distribution are set as (1, 2, 1, 2, 1) and ($\pi$, $\pi/2$, 0, $\pi/2$, $\pi$). The simulated patterns and experimental patterns of input light are presented in Fig. 6(c). Figures 6(d) and 6(e) show the measured OAM power spectra without and with reference light respectively. The measurement errors are 0.0768 and 0.0681 respectively. In the case, the results in Fig. 6(d) indicate the inherent noise from the cross terms and the results in Fig. 6(e) are the OAM power spectrum with the same inherent noise. With subtraction of twice measurements, the inherent noise can be greatly suppressed as shown in Fig. 6(f) and the measurement error reduces to 0.0320. In the case, the errors are mainly from the uneven conversion efficiency and the phase distribution cannot be decided. The results are then calibrated by normalization with the reference data in Figs. 6(a) and (b). The measurement error (about 0.0063) becomes smaller further. The final power distribution and phase distribution of OAM modes are presented in Figs. 6(g) and (h) respectively, which agree well with the theoretical calculations. In Fig, 6(h), the phase of the OAM modes whose power is very small is amended to null. Figure 7 demonstrates a more general and complex measurement with calibration, where the unknown amplitude distribution is set as a triangular function and the phase is set irregularly. The simulated patterns and experimental patterns of input light are presented in Fig. 7(a). The final power distribution and phase distribution of OAM modes are presented in Figs. 7(b) and (c) respectively. The measurement error is about 0.0055. We can see that the measured power distribution matches well the theoretical calculations. Although there are some apparent errors in the phase measurement, it is acceptable because the corresponding OAM modes have quite low power.

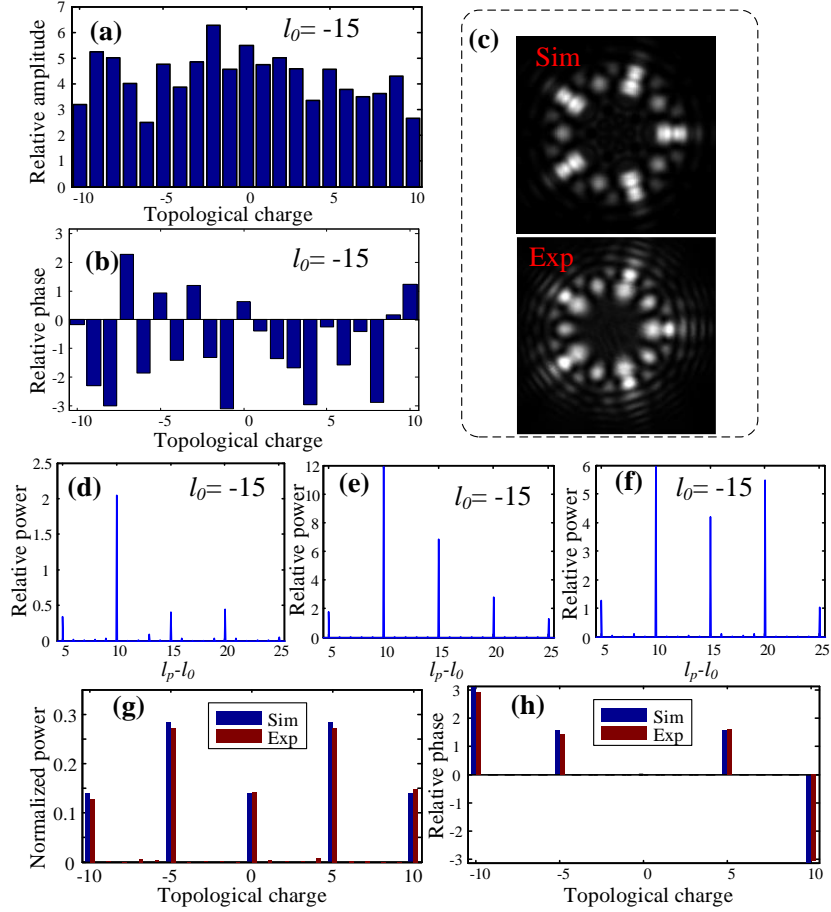

Figure 6. Process of Calibrating OAM complex spectrum for a general spinning object. (a) The amplitude distribution and (b) the phase distribution of beating signals between every input OAM modes with the reference mode. The data are acquired via measuring the complex amplitude of corresponding beating signal one by one. (c) The simulated patterns and experimental patterns of input light. (d) The measured OAM power spectrum without reference light. (e) The OAM power spectrum with reference light. (f) The OAM power spectrum of twice measurements. (g, h) The final power distribution and phase distribution of OAM modes compared to the theoretical calculations.

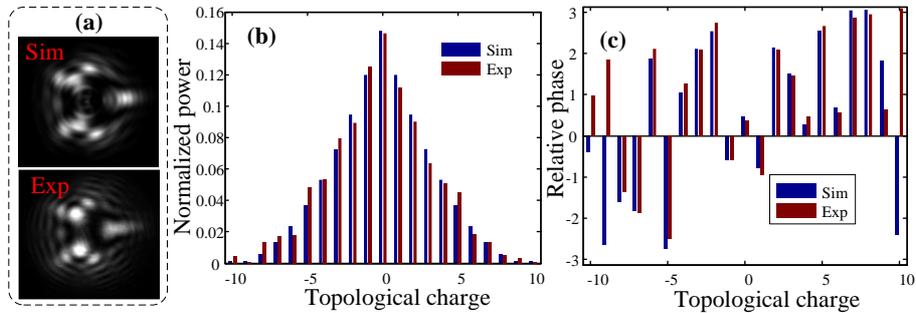

Figure 7. Experimental data after calibration. The amplitude distribution of input OAM modes is set as a triangle. (a) The simulated patterns and experimental patterns of input light. (b) The final power distribution and (c) the phase distribution of OAM modes compared to the theoretical calculations.

## Discussion

In the above, we design an experiment to demonstrate our scheme at the proof-of-concept level. We use an SLM (SLM1) to generate the input light to be measured and the reference light simultaneously. Note that according to the theory of rotational Doppler Effect, the spinning object is not necessarily really rotated. Equation (5) reveals the rotation of object is equivalent to the rotation of surface modulated function of fixed object. Therefore, another SLM (SLM2) is employed to emulate the spinning object. We scan the azimuth by rotating the pattern loaded in SLM2 and record the collected intensity by a CCD. Since the frame rate of SLM is about 60 Hz, it will take a few seconds to complete the measurement. On the other hand, a real spinning object will increase the requirement of mechanical performance or introduce mechanical vibrations resulting in unstable working state. In a real-world scenario, we can use a digital mirror device (DMD) instead of SLM2. The DMD has a similar function to the SLM but with a higher frame rate up to 20 kHz, and it is fast enough for practical application. Besides, our configuration works only for a single wavelength, so we can introduce the reference light from an additional tunable laser in real deployment, similar to the local oscillator in coherent optical communication systems[33].

Our scheme can be also extended to measure the mode distribution in few-mode fiber only by mapping the OAM modes to the eigenmodes of fiber. In most cases, only a small number of eigenmodes are used as independent channels, such as spatial modes (i.e., LP01, LP11$a$, LP11$b$, LP21$a$ and LP21$b$, where LP represents linear polarization modes, $a$ and $b$ represent the even modes and the old modes respectively) in few-mode fiber. These modes have differences only in the azimuthal mode index which is similar to OAM modes. In fact, the OAM modes are the orthogonal basis of Fourier expansion by complex exponential series, and the LP modes are the orthogonal basis of Fourier expansion by trigonometric series. These two bases can be converted to each other. So we just need measure the OAM complex spectrum firstly and then map it to the complex spectrum based on LP modes. The ability of measuring the mode distribution in fiber shows great applications in optical fiber communication.

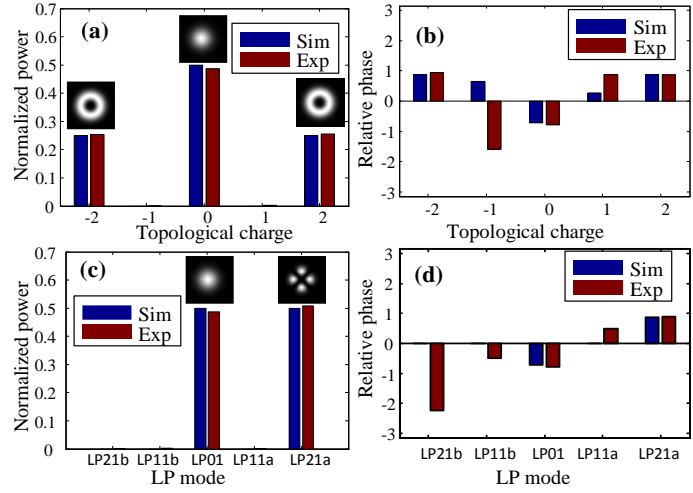

Figure 8. Experimental results on (a, b) OAM basis and (c, d) LP basis. The inset patterns are the intensity distributions of corresponding modes.

Figure 8 presents a simple experimental demonstration on the ability of measuring the LP mode distribution. The input light only contains two modes, LP01 and LP21a. They have the same power with a $\pi/2$ phase difference. According to our scheme, we firstly measure the OAM complex spectrum as shown in Figs. 8(a) and (b). And then these OAM modes are expressed on the LP basis. The final results presented in Figs. 8(c) and (d) are accurate, proving the ability of LP mode measurement. Note that the measured phase of LP21*b*, LP11*b* and LP11*a* can be revised to null because those amplitudes are null.

# CONCLUSIONS

In summary, we demonstrate a prototype of OAM complex spectrum analyzer based on rotational Doppler Effect. The system consists of unknown input OAM beams, a strong reference light, a spinning reflective object, a mode filter and a photodetector. The OAM power spectrum is measured with inherent noise via initial measurement by introducing a strong reference light. The noise is mainly from the cross terms of input OAM modes and can be reduced by increasing the power ratio between the reference light and the input light. We further demonstrate that the noise can be eliminated by subtraction of the initial measurement and another constant-noise measurement. Finally, we give an example to calibrate the OAM power spectrum for a general spinning surface, whose modulation function has an uneven distribution in Fourier expansion along azimuthal direction. And through calibration, the phase distribution can be acquired. Similar to the optical spectrum analyzer, the ability to measure the OAM complex spectrum of light has important applications for future OAM-based optical communication systems, few-mode fiber systems, and optical quantum systems.

# ACKNOWLEDGMENTS

This work was partially supported by the National Basic Research Program of China (Grant No. 2011CB301704), the Program for New Century Excellent Talents in Ministry of Education of China (Grant No. NCET-11-0168), a Foundation for the Author of National Excellent Doctoral Dissertation of China (Grant No. 201139), the National Natural Science Foundation of China (Grant No. 11174096, 11374008, 11534008 and 61475052), Foundation for Innovative Research Groups of the Natural Science Foundation of Hubei Province (Grant No. 2014CFA004).

# AUTHOR CONTRIBUTIONS

H.L.Z. and D.Z.F. contributed equally to this paper. H.L.Z. proposed the study. H.L.Z., D.Z.F. and D.X.C. carried out the experiment. P.Z. supervised the experiment. H.L.Z. and J.J.D. analyzed the results and wrote the manuscript. J.J.D., P.Z. F.L.L and X.L.Z. supervised the project and edited the manuscript. All authors discussed the results and commented on the manuscript.

## ADDITIONAL INFORMATION

The authors declare no competing financial interests. Correspondence and requests for materials should be addressed to J.J.D. and P.Z.